\documentclass{aastex}
 \newcommand{\RXTE}{\emph{Rossi X-ray Timing Explorer}}
 \newcommand{\rxte}{\emph{RXTE}}
 \newcommand{\ASM}{All Sky Monitor}
 \newcommand{\asm}{ASM}
 \newcommand{\PCA}{Proportional Counter Array}
 \newcommand{\pca}{PCA}
 \newcommand{\pcu}{PCU}
 \newcommand{\pcus}{PCUs}
 \newcommand{\HEXTE}{High Energy X-ray Timing Experiment}
 \newcommand{\hexte}{HEXTE}

 \newcommand{\sax}{\emph{Beppo}SAX}
 \newcommand{\ginga}{\emph{Ginga}}

 \newcommand{\asca}{\emph{ASCA}}

 \newcommand{\her}{Her X-1}
 \newcommand{\flc}{FLC}

 \newcommand{\hecut}{HECUT}
 \newcommand{\xray}{X-ray}
 \newcommand{\xrays}{X-rays}
 \newcommand{\ALS}{Anomalous Low State}
 \newcommand{\als}{ALS}
 \newcommand{\dof}{DOF}

 \newcommand{\pcmsq}{$\mathrm{cm^{-2}}$}
 
 \newcommand{\flux}{$\mathrm{ergs\,cm^{-2}\,s^{-1}}$}

 \newcommand{\porb}{$P_{\rm{orb}}$}
 
 \newcommand{\wsim}{$\sim$}
 \newcommand{\wtimes}{$\times$}
 
 \newcommand{\wchi}{$\chi^{2}$}
 \newcommand{\wchired}{$\chi^{2}_{\rm{red}}$}
 \newcommand{\wcalf}{$\mathcal{F}$}
 \newcommand{\wgamma}{$\Gamma$}
 \newcommand{\we}{\begin{equation}}
 \newcommand{\ew}{\end{equation}}

 \newcommand{\nhu}{\mathrm{N}_{\mathrm{H}}\ (10^{22}\,\mathrm{cm^2})}
 \newcommand{\nh}{\mathrm{N}_{\mathrm{H}}}
 \newcommand{\ec}{E_{\mathrm{c}}}
 \newcommand{\ef}{E_{\mathrm{f}}}
 \newcommand{\gosc}{\gamma_{\rm pulse}}
 \newcommand{\gosca}{\gamma_{\rm pulse,A}}
 \newcommand{\goscb}{\gamma_{\rm pulse,B}}
 \newcommand{\goscc}{\gamma_{\rm pulse,C}}
 \newcommand{\goscd}{\gamma_{\rm pulse,D}}

 \newcommand{\hecuteq}{ \mathrm{HECUT}(E) = A\ E^{-\Gamma} \left\{
    \begin{array}{ll}
       1                            & \mbox{$(E<\ec)$} \\
       \mbox{$\mathrm{e}^{-(E-\ec)/\ef}$} & \mbox{$(E>\ec)$}
    \end{array}\right.}

 \newcommand{\modeleq}{F_{\mathrm{pa}}(E) = (1\ +\ f \times
    e^{-\sigma(E)\nh}) (\mathrm{GAUSS}\ +\ \mathrm{HECUT}) }

 \newcommand{\conteq}{ F(E) = \mathrm{e}^{-\sigma(E)\nh} (\mathrm{GAUSS}\
    +\ \mathrm{HECUT})}

\begin{document}

%% ********************************************************************
%% ** The Header                                                     **
%% ********************************************************************
\title{The 1999 Hercules X-1 Anomalous Low State}
\author{W. Coburn\altaffilmark{1}, W. A. Heindl\altaffilmark{1},
J. Wilms\altaffilmark{2}, D. E. Gruber\altaffilmark{1}, R.
Staubert\altaffilmark{2}, R. E. Rothschild\altaffilmark{1}, K. A.
Postnov\altaffilmark{3}, N. Shakura\altaffilmark{4}, P.
Risse\altaffilmark{2}, I. Kreykenbohm\altaffilmark{2}, M. R.
Pelling\altaffilmark{1}}

\altaffiltext{1}{Center for Astrophysics and Space Sciences, Code
0424, University of California at San Diego, La Jolla, CA,
92093-0424, USA}

\altaffiltext{2}{Institut f\"{u}r Astronomie und Astrophysik,
Astronomie, University of T\"{u}bingen, Waldh\"{a}user Strasse 64,
D-72076 T\"{u}bingen, Germany}

\altaffiltext{3}{Faculty of Physics, Moscow State University,
119899 Moscow, Russia}

\altaffiltext{4}{Sternberg Astronomical Institute, 119899 Moscow,
Russia}

%% ********************************************************************
%% ** The Abstract                                                   **
%% ********************************************************************
\begin{abstract}
A failed main-on in the 35\,d cycle of \her\ was observed with the
\RXTE\ (\rxte) on 1999 April 26. Exceptions to the normal 35\,d
cycle have been seen only twice before; in 1983 and again 1993. We
present timing and spectral results of this latest \ALS\ (\als)
along with comparisons to the main-on and normal low states.
Pulsations were observed in the 3-18\,keV band with a fractional
RMS variation of ($0.037 \pm 0.003$).  Spectral analysis indicates
that the \als\ spectrum has the same shape as the main-on but is
modified by heavy absorption and scattering.  We find that 70\% of
the observed emission has passed through a cold absorber
($\nh=5.0\times10^{23}$\,\pcmsq).  This partially absorbing
spectral fit can be applied to the normal low state with similar
results. We find that the \als\ observations may be interpreted as
a decrease in inclination of the accretion disk causing the
central \xray\ source to be obscured over the entire 35\,d cycle.
\end{abstract}

%% ********************************************************************
%% ** Introduction                                                   **
%% ********************************************************************
\section{Introduction}

\her\ is a well known persistent accreting \xray\ pulsar. It is
characterized by a 1.24\,s pulse period \citep{tan72} and a 1.7\,d
orbit \citep{dee81} around the A/F star HZ Her
\citep{dox73,got91}. \her\ also exhibits a 35\,d intensity cycle
that is generally believed to be due to the precession of a
tilted, warped accretion disk viewed nearly edge on, which
periodically obscures \xrays\ from the central neutron star. For a
review of earlier models of \her, see \citet{pri87}.  Recently,
more detailed models of the physical cause of the warping have
been advanced \citep{sha99,pri96,sch94}.

During the 35\,d cycle, \her\ normally has a \wsim 10\,d main-on
state and a \wsim 5\,d short-on state, each separated by \wsim
10\,d low states. The peak count rate in the \RXTE\ \ASM\
(\rxte/\asm, 1.5-12\,keV) for the short-on is usually \wsim 35\%
that of the main-on while the low state is \wsim 3\%
\citep{sco99}.  Eclipses are seen at the 1.7\,d orbital period, as
are regular pre-eclipse absorption dips every 1.62 days (the
1.7\,d and 35\,d beat period; \citet{gia73,cro80}), and anomalous
dips shortly after the main-on turn-on. The \xray\ absorption dips
are discussed in more detail by \citet{ste99}, \citet{sha98},
\citet{lea97}, and \citet{rey95}.

Exceptions to this usual behavior were observed in 1983 June
\citep{par85}, in 1993 August \citep{vrt94,vrt96} and most
recently in 1999 April \citep{lev99,par99}. The 1999 April \als\
has continued into the year 2000. The \ALS\ (\als) has been
defined by \citet{vrt96} as an unexpected and substantial drop in
\xray\ flux that persists for more than two 35\,d cycles, with no
substantial change in absorbing column density, little or no
change in UV and optical fluxes, an increase in pulse period, and
no pulsed emission above \wsim 1.0 keV. The present data show that
this definition of Vrtilek needs modification. The emission is
found to be pulsed, but with a factor of \wsim14 reduction
fractional RMS variation from the main-on.  Also, we find evidence
for a highly absorbed component in the \xray\ spectrum. In the
1983 \als\ the neutron star was believed to be steadily obscured
due to a temporary change in the accretion disk \citep{par85}.

The beginning of the 1999 \als\ is very similar to the 1983 \als\
in that the source did not turn on at the date predicted by the
mean 35\,d ephemeris, while in 1993 the \xray\ flux suddenly
decreased during the middle of the main-on \citep{vrt94}.  Optical
observations made during the 1983 \als\ \citep{del83,mir86}
indicated that HZ Her remained heated by \xray\ emission from the
central source, implying that accretion was still proceeding
normally onto the neutron star surface. Similarly, in 1993 the
optical and UV fluxes continued to show the 1.7\,d modulation
\citep{vrt94}, but with a reduction in UV flux observed around
eclipse. UBV photometry of HZ Her during the 1999 \als\ revealed a
light curve similar to the one from 1983 \citep{lyu99}.

On 1999 March 23, the first deviation from the normal behavior in
three years of \asm\ observations of the \her\ main-on state was
detected. Typical peak intensities are 40-100\,mCrab in the \asm\
but the upper limit for this main-on was \wsim15\,mCrab
\citep{lev99}.  On 1999 April 26, during the second failed
main-on, we carried out an \rxte\ observation of \her\ to
investigate the nature of this \als.  We discuss spectral and
temporal analyses of this \rxte\ observation which support the
idea that the \xray\ source is continuously screened by the
accretion disk during an \als.

%% ********************************************************************
%% ** Observations and Analysis                                      **
%% ********************************************************************
\section{Observations and Analysis}

The \als\ observation was made using the \PCA\ (\pca)
\citep{jah96} and the \HEXTE\ (\hexte) \citep{rot98} on board the
\rxte.  The source was observed on 4 consecutive \rxte\ orbits
spanning an elapsed time of 20\,ks centered at MJD 51294.52. We
compared this observation to archival \rxte\ data taken during the
normal low state and normal main-on.  See Table~\ref{table:obs}
for a summary of the observations. During this discussion we refer
to the \als\ observation unless specifically noted otherwise.

Since main-ons were not observed, we cannot be certain that the
observation fell during the second failed main-on of the current
\als. On average, main-ons occur at a period of 20.5 times the
orbital period (\porb), but are also observed at 20.0 and
21.0\wtimes\porb\ as well. If, after the last observed main-on,
both of the expected turn-ons occurred at the late value of
21\wtimes\porb, then the source was observed during a low state
approximately one half of a binary orbit before turn-on.  If, as
is more likely, either of the unobserved turn-ons were at 20.0 or
20.5\wtimes\porb, then the source was observed during a failed
main-on.

\subsection{Timing Analysis\label{subsec:timing}}

\begin{figure}[ht]
\centerline{\includegraphics{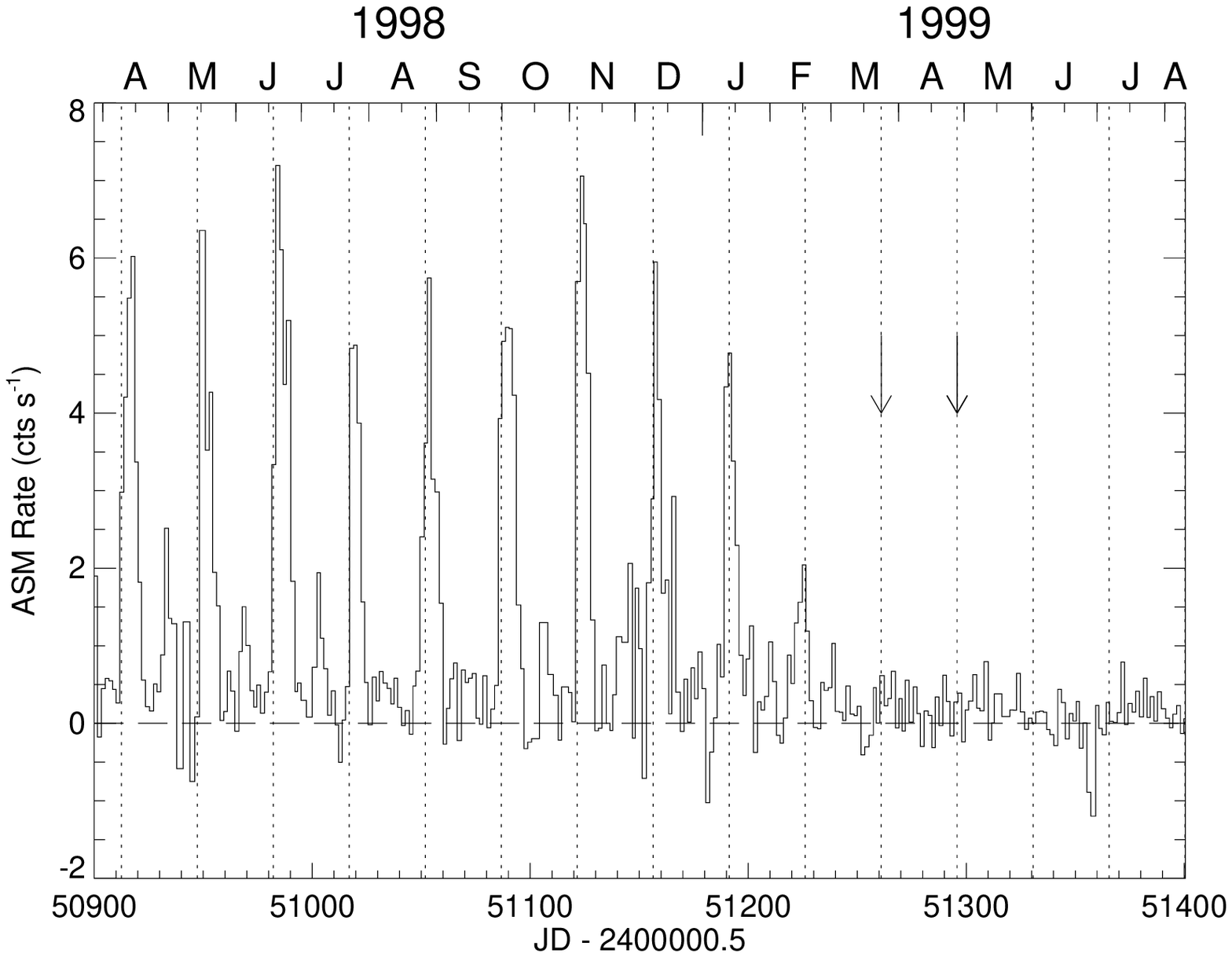}}
\figcaption{\label{fig:asm_lc} The \rxte/\asm\ light curve of
\her. The dotted lines indicate main-on turn-ons using the
ephemeris of \citet{sco99}. The bins are the average count rate in
each 1.7\,d orbit of the system. The start of the \als\ can be
clearly seen at the first arrow. Our observation occurred at the
second. The light curve also shows evidence for a gradual change
into the \ALS.}
\end{figure}

Fig.~\ref{fig:asm_lc} shows the \asm\ light curve for the last ten
35\,d cycles prior to the failed main-on turn-on. Each bin in the
light curve is the average counting rate in a single 1.7\,d orbit
of the pulsar with eclipse intervals excluded using the ephemeris
of \citet{dee91}. The last three main-ons have a steadily
decreasing count rate. Thus, the transition to this \ALS\ was a
gradual process occurring over several cycles.

\begin{figure}[ht]
\centerline{\includegraphics[]{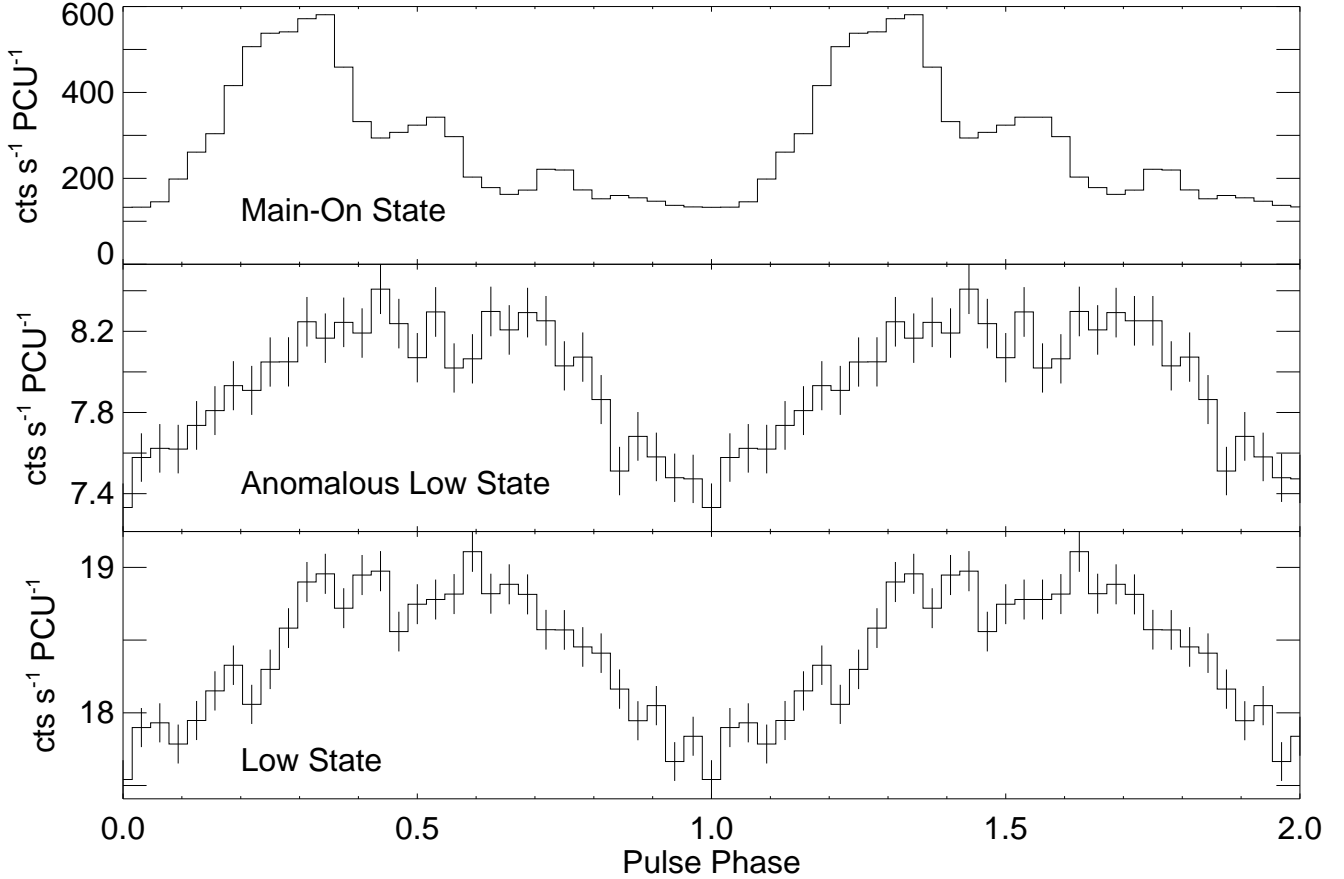}}
\figcaption{\label{fig:pca_flc} The \pca\ folded light curves of
\her\ from 3-18\,keV.  Top: From the peak of the normal main-on.
Middle: From the 1999 \ALS. Bottom: During the low state, just
before turn-on. The phase alignment of the three \flc s is
arbitrary. Note the offset of the $y$ axes in the bottom two
panels. The \als\ folded light curve is much shallower and broader
than what is seen in during the main-on, but very similar to that
of the low state. Despite the fact that there is less overall flux
in the \als\ pulse, the fractional RMS variation is slightly
larger than that of the low state (see
Sect.~\ref{subsec:timing}).}
\end{figure}

We generated the 3-18\,keV \pca\ folded light curves (\flc s)
(Fig.~\ref{fig:pca_flc}) by first creating a background subtracted
light curve for the observation. The background count rate per bin
was estimated using the count rates from the \pca\ modeled
background spectrum over the same channel range. We then corrected
the photon arrival times to the solar system and pulsar binary
system barycenters and performed a \wchi\ period search. The
period at epoch MJD 51294.5 was 1.2377485(3)\,s. We determined the
error by folding the data on nearby periods and calculating the
\wchi\ statistic with respect to the \flc\ of our best period.
This best period is consistent with the 1.237747(2) s period found
by \citet{par99} with the \sax\ satellite during a 1999 July 8-10
observation. Periods were found using the same method for the
main-on and normal low state data.

Compared to the main-on, the \als\ folded light curve is much
shallower and broader (Fig.~\ref{fig:pca_flc}). But when compared
to the low state, the \als\ \flc\ is similar in shape but with
reduced flux. This indicates that the mechanism for creating the
\als\ might be similar to that of the low state, namely
occultation by the accretion disk, but also accompanied by a
decrease in mass accretion rate. In and of itself this is
inconclusive, however, since the pulse shape of \her\ is known to
vary considerably with the 35\,d cycle \citep{sco00,dee98}. The
nearly sinusoidal pulse shapes seen in the low and \als\ states in
Fig.~\ref{fig:pca_flc} have been observed before, but always near
the end of the main-on and secondary high state and not at
turn-on.

\begin{figure}[ht]
\centerline{\includegraphics[]{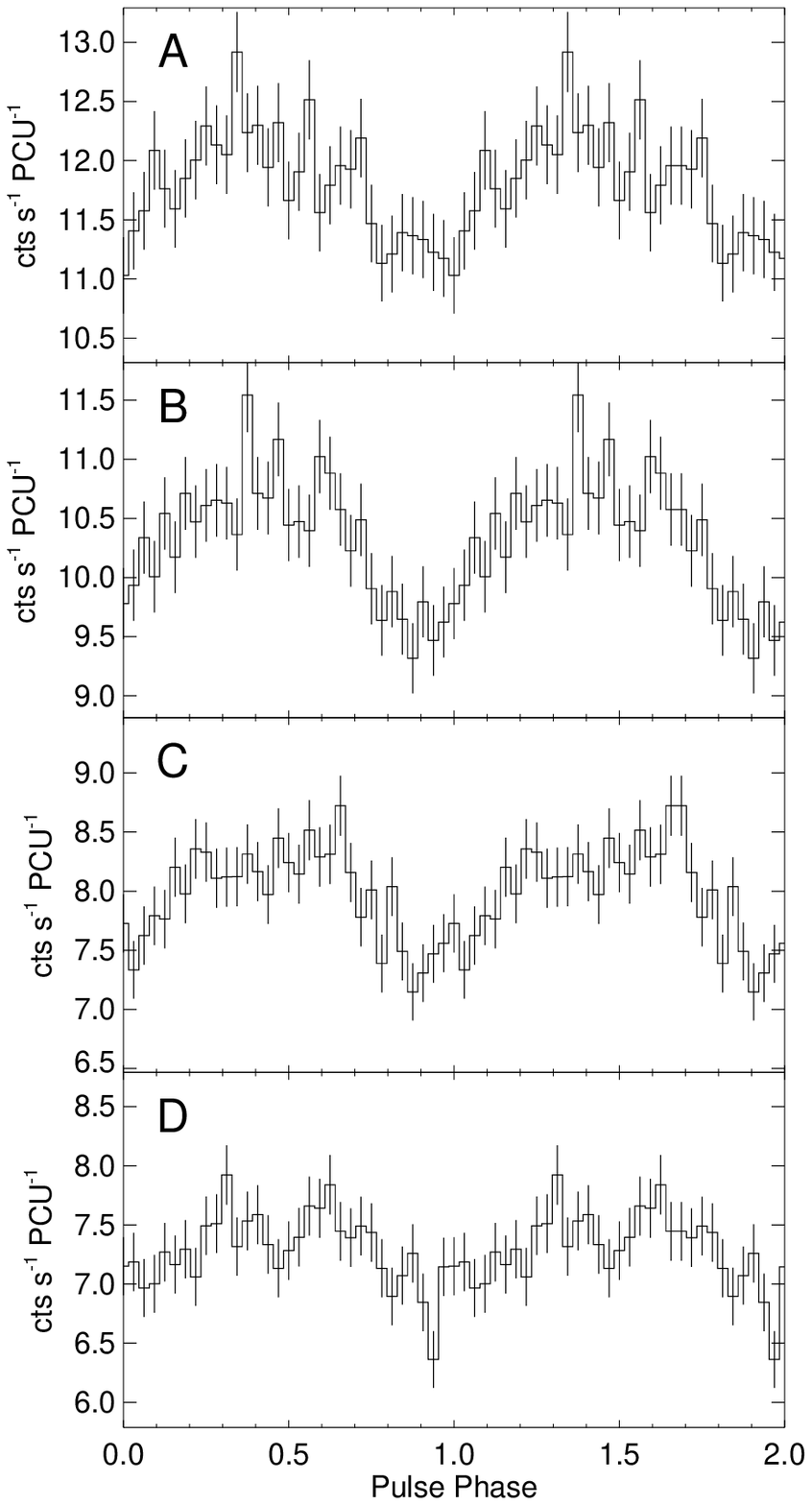}}
\figcaption{\label{fig:pca_orbitflc} The \rxte/\pca\ 3-18\,keV
folded light curve for four intervals as labeled in
Fig.~\ref{fig:pca_lc}. Pulsations are clearly seen in all four
panels. The fractional RMS variation (see
Sect.~\ref{subsec:timing}) of the four intervals are $\gosca =
0.038 \pm 0.005$, $\goscb = 0.049 \pm 0.005$, $\goscc = 0.049 \pm
0.006$, $\goscd = 0.041 \pm 0.006$.}
\end{figure}

A useful measure of the amount of pulsations is the fractional RMS
variation $\gosc \equiv <(\Delta I)^{2}>^{1/2}/<I>$, where $\Delta
I \equiv I - <I>$ and the angle brackets indicate averaging with
respect to the pulse phase. In this definition the Fourier power
at the rotation frequency seen is proportional to $\gosc^{2}$. For
the \als\ state $\gosc = 0.037 \pm 0.003$ in the 3-18\,keV band.
This is much lower than the value of $\gosc = 0.5152 \pm 0.0003$
observed in the main-on, but similar to the low state value of
$\gosc = 0.024 \pm 0.001$. We also found that the fractional RMS
variation remained nearly constant throughout our observation
(Fig.~\ref{fig:pca_orbitflc}). The inferred 1.2-37.2\,keV upper
limit for pulsations in the low state as seen with \ginga\ is
$\gosc \leq 0.017$ \citep{mih91}, which is less than what we found
with the \rxte\ at the same orbital phase. This implies that the
fractional RMS variation from one low state to another is
variable. The similarity of the \als\ and low state $\gosc$ would
seem to imply that they are both due to similar mechanisms,
specifically occultation by the accretion disk.

\begin{figure}[ht]
\centerline{\includegraphics[]{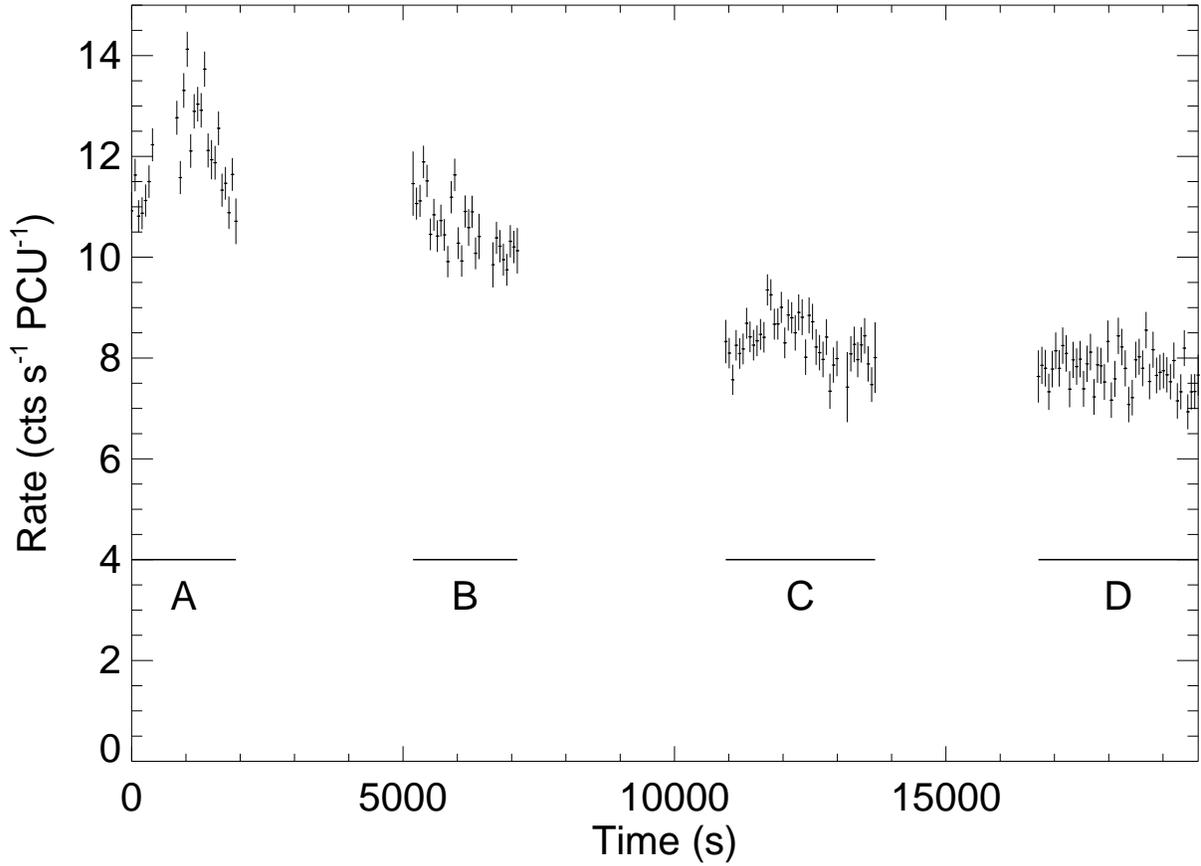}}
\figcaption{\label{fig:pca_lc} The \rxte/\pca\ 3-18\,keV light
curve, with 64\,s time resolution, beginning at MJD 51294.4 (1999
April 26 9:40UTC). The gaps between the data are due to earth
occultations and SAA passages. Although the flux decreased during
the observation, the spectral shape of each segment remained
unchanged. The \flc s for the intervals labeled A-D are shown in
Fig.~\ref{fig:pca_orbitflc}}
\end{figure}

The \pca\ light curve (Fig.~\ref{fig:pca_lc}) shows a gradual drop
of flux by 40\% through the observation. The binary orbital phase
of the observation is 0.38-0.52, too early for the decrease to be
due to a pre-eclipse dip. We compared the spectra from the four
\rxte\ orbits and found that the spectral shape and fractional RMS
variation remained the same. Fig.~\ref{fig:pca_orbitflc} shows the
pulse shape over the course of the observation. A similar
reduction in flux was observed in this \als\ with the \sax\
\citep{par99}. There, at orbital phase \wsim0.3 and most
strikingly in the 1.8-10\,keV band, the flux began to steadily
decrease until eclipse, after which it steadily increased,
recovering again at orbital phase \wsim1.3. Something similar may
have been seen with \asca\ \citep{vrt94} in 1993, but the
observation did not start until orbital phase \wsim0.7.

\subsection{Spectral Analysis}

We generated \pca\ and \hexte\ spectra from the data using the
FTOOLS 4.2. To account for uncertainties in the \pca\ epoch 4
preliminary response matrix (1999 March 30) and background models,
we added 2\% energy independent systematic errors to the \pca\
data. This 2\% is typical of the errors in fits to the Crab
Nebula/Pulsar during the same epoch. We only accumulated data from
the top layer of \pcus\ 0-3 in the \pca. \pcu\ 4 was off during
the observation.

We fit the three spectra (\als, main-on, and normal low state) to
standard accreting \xray\ pulsar continuum models (high energy
cut-off power law (\hecut), Fermi-Dirac cut-off power law, and
negative-positive exponential \citep{mih95,kre99}. We restrict our
discussion to fits made with the \hecut\ model since the \hecut\
has been used historically to describe \her\ and the statistics of
the observation do not require one of the other continua. The
model is of the form \we\label{eq:conteq}\conteq\ew where
$\sigma(E)$ is the photoelectric absorption cross section of the
interstellar medium \citep{mor83}, $\nh$ is the column density,
and the Gaussian represents an FeK line. The \hecut\ is
\we\label{eq:hecut}\hecuteq\ew where \wgamma\ is the photon
powerlaw index, and $\ec$ and $\ef$ the cutoff and folding
energies respectively. In previous work for the main-on state it
has found that the neutral column density was less than
$\sim10^{20}$\,\pcmsq\ \citep{dal98}, so we fixed this value to
zero when doing our main-on spectral fits.

The statistics of our observation of the \als\ spectrum do not
require the well known \her\ cyclotron resonance scattering
feature (CRSF) at \wsim40\,keV \citep{gru98,dal98}. A CRSF whose
fit parameters were constrained to be identical to those in the
main-on fit did not significantly improve the fits. Rather than
compare dissimilar models for the continuum we also fit the
main-on spectrum without a CRSF. The primary difference between
the main-on continuum shapes with and without a CRSF is the
folding energy, which is important only at higher energies
($\gtrsim20$\,keV) where the statistics of the \als\ data become
poor. The poor \wchired\ of the high state fit
(Table~\ref{table:fits}) is due to the lack of a CRSF feature in
the model.

Initially we modeled the \als\ assuming only an \hecut\ continuum
model with an FeK line and photoelectric absorption
(Eq.~\ref{eq:conteq}, see Table~\ref{table:fits}). The best fit
required no low energy absorption ($\nh < 4 \times
10^{21}$\,\pcmsq), a photon index of $0.37\pm0.04$, and resulted
in a \wchired of 1.12 for 114 degrees of freedom (\dof). This
spectral index is very different from the \wgamma\wsim1 that has
been seen historically during the main-on. Because the spectral
index fit with Eq.~\ref{eq:conteq} is different from what is seen
in the main-on, we decided to test if there was a fit that left
the intrinsic, main-on spectral shape unchanged.  A change in the
accretion disk structure could lead to a combination absorption
from the disk and scattering from a hot corona (physically
distinct but numerically very similar is a partial-covering
model). Using a similar analysis to that of \citet{mih91} and
\citet{par99} we fit the \als\ spectrum with a partially absorbing
column of the form \we\label{eq:model}\modeleq\ew We define a
covering factor \wcalf\ as $f/(1+f)$, which is the fraction of
observed flux coming to us through the absorbing medium.

\begin{figure}[ht]
\centerline{\includegraphics[]{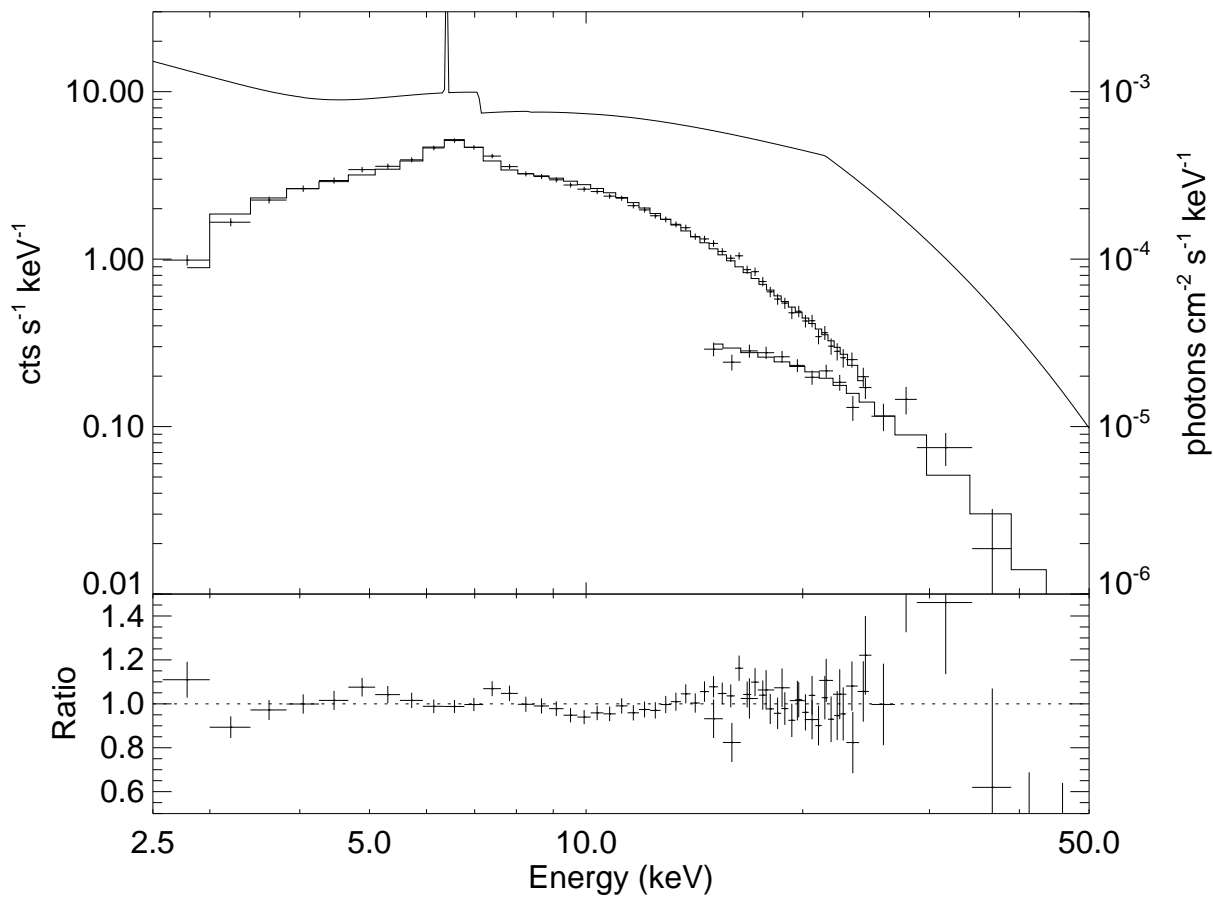}}
\figcaption{\label{fig:spec_als} Top: The joint \rxte\
\pca/\hexte\ spectrum of \her\ during a failed main-on in the 1999
\als\ (crosses).  The fit with Eq.~\ref{eq:model} (the partial
absorption/transmission model, the shape of the \hecut\
constrained to that of the main-on) is shown as histograms, along
with the model photon spectrum (smooth curve). Bottom: The ratio
of the data to the fit model.}
\end{figure}

\begin{figure}[ht]
\centerline{\includegraphics[]{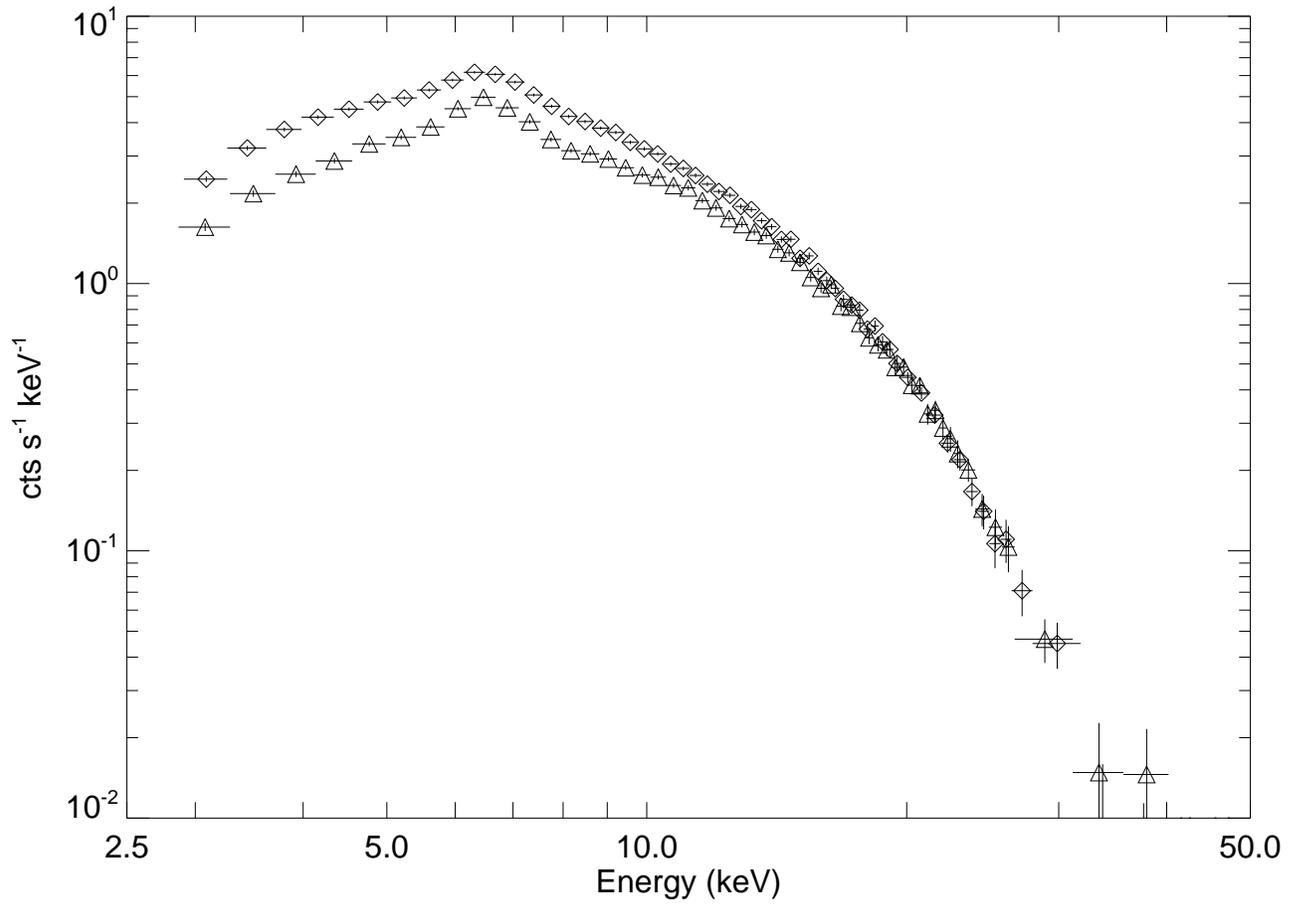}}
\figcaption{\label{fig:spec_low} \pca\ counts spectra for the low
state (diamonds) and the \ALS\ (triangles). The overall shape of
the \als\ spectrum is similar to that of the normal low state, but
with an overall reduction in flux at lower energies (below \wsim
15\,keV) and an enhanced Fe-K line. }
\end{figure}

With the exception of the normalization, the best fit \hecut\
parameters using Eq.~\ref{eq:model} are consistent with those of
the main-on (see Fig.~\ref{fig:spec_als} for the \pca/\hexte\
counts spectra and best fit model, Table~\ref{table:fits} for fit
parameters). Thus, what is seen in the \als\ is a highly absorbed
and heavily scattered main-on spectrum. The small improvement in
the fit from the simple \hecut\ continuum (to \wchi/113=1.08) is
not statistically significant, but the model makes more physical
sense. When the \hecut\ was constrained to be the same shape as
the main-on, the \wchi/117 is 1.07 with $\nh = (5.0 \pm 0.3)\times
10^{23}$\,\pcmsq\ and $\mathcal{F} = 0.67 \pm 0.04$. In addition,
the \als\ and normal low state spectra are similar (see
Fig.~\ref{fig:spec_low}). The low state fit with
Eq.~\ref{eq:model} and the main-on \hecut\ continuum shape gives
$\nh = (4.8 \pm 0.4)\times 10^{23}$\,\pcmsq\ and $\mathcal{F} =
0.54 \pm 0.02$. This is a slightly lower column density and higher
covering factor than what was seen by \citet{mih91}, implying that
low state is somewhat variable. The \als\ spectrum observed by
\sax\ was found to be indistinguishable from that observed during
the normal low state \citep{par99}. So, the spectra are consistent
with the same mechanism causing both the normal low and anomalous
low states.

The FeK line is very prominent in the \als\ spectrum, with an
equivalent width of $0.7\pm0.1$\,keV.  This is also similar to
what has been observed in the low state.  For example
\citet{mih91} reported $1.0\pm0.1$\,keV, and we find
$0.5\pm0.1$\,keV in the low state (Table ~\ref{table:fits}). We
also fit the data with a superposition of two iron lines, at
6.4\,keV and 6.7\,keV, to search for evidence of an ionized iron
line component. There was no improvement with a double line fit,
and the best fit was for a single line at $6.4\pm0.1$\,keV.

%% ********************************************************************
%% ** Discussion                                                     **
%% ********************************************************************
\section{Discussion}

The measured pulse period of 1.2377485(3)\,s from the first \rxte\
observation during the 1999 \als\ establishes a rapid spin-down of
the neutron star. The period is about 25\,$\mu$s longer than what
would have been expected if the \her\ general spin-up trend
\citep{bil97} had continued. This is also the largest deviation
from the general spin-up among the three observed anomalous low
states: for the \als\ in 1983 \citep{par85} the deviation was
about 20\,$\mu$s and in 1993 \citep{vrt94} it was about
12\,$\mu$s. In standard accretion torque theory \citep{gho79}, a
reduction in mass accretion (with an accompanied reduction in
\xray\ luminosity) leads to a reduced gas pressure in the
accretion disk. This, in turn, leads to an increased
magnetospheric radius and a change in the accretion torque that
causes the neutron star to spin-down. It is expected that a
reduction in mass accretion would also cause a reduction in the
inclination of the accretion disk, either by changing the torque
of the accretion stream on the disk \citep{sha99} and/or by the
radiative torque from a coronal wind resulting from the \xray\
heating of the accretion disk \citep{sch94,sch96}. If the
reduction in inclination angle of the disk were large enough it
would continuously screen the central \xray\ source from our line
of sight. We also note that, before the onset of the current \als,
the \xray\ flux continuously dropped (Fig.~\ref{fig:asm_lc}),
possibly due to both a reduced mass accretion rate and increased
obscuration by the disk.

Recent optical observations \citep{lyu99} show that, as in
previous \als s, HZ Her continues to be heated by the \xray\
source \citep{del83,mir86} at a level similar to what is normal.
So, \xrays\ are still being produced and mass accretion onto the
surface of the neutron star must therefore be continuing without
any great change.  If there were no occultation by the disk (i.e.,
if the main-on geometry were unchanged), then to account for the
factor of 17 drop in flux at high energies ($\gtrsim20$\,keV) by
scattering alone would require a Thomson optical depth of \wsim3.
Whether or not this could explain the order of magnitude change in
the fractional RMS variation is unclear. The calculations by
\citet{bra87} and \citet{kyl87} assume Thomson optical depths
$\tau \gtrsim 5$ and a relatively simple input pulse shapes. The
complexity of the \her\ pulse shape during the main-on, with the
additional complications of viewing angles and corona size, makes
it difficult to directly apply their results. It is clear,
however, that scattering alone cannot account for the change in
spectral shape. A more reasonable explanation is that the the
neutron star is occulted along the line of sight by the accretion
disk, just as it is known to be during the low state. This change
from the normal 35\,d cycle could be caused by either a change in
the disk inclination or the amount of warp of the disk.

Optical observations over a binary orbit \citep{lyu99} also show
that the secondary minimum near the peak of the light curve, which
is due to the accretion disk occulting the heated surface of HZ
Her, has almost disappeared. This also provides evidence that the
accretion disk has become less inclined with respect to the
observer's line of sight. The decrease in the disk inclination
should be accompanied by a decrease in the heating of the star's
surface near the inner Lagrangian point. This can explain why,
away from maximum brightness, the \als\ optical light curve is
lower than the average optical light curve. In turn, the decrease
in heating near the inner Lagrangian point leads to a decrease in
the accretion rate onto the neutron star.  In the model of
\citet{sha99}, the equilibrium inclination angle of outer parts of
the accretion disk to the orbital plane is determined by the joint
action of accreting streams and tidal forces, so the decrease in
accretion rate leads to a smaller inclination angle.  A decrease
in inclination angle of the accretion disk, moving it more into
the line of sight, also explains the continued occultation of the
neutron star and our \als\ observations.

The 2.5-50\,keV \rxte\ spectrum of the \als\ and normal low states
are similar (see Fig.~\ref{fig:spec_low} and
Table~\ref{table:fits}). Above \wsim15\,keV the two spectra are
nearly identical, while at lower energies the \als\ spectrum has
less flux and a more pronounced Fe-K line. Fits using the main-on
\hecut\ continuum shape (Eq.~\ref{eq:model}) and a partially
covering screen of absorbing matter gave similar results for both
the normal and anomalous low states. A similar analysis of the low
state was done by \citet{mih91}. There the inferred covering
factor was $\mathcal{F}=0.55\pm0.03$, which is consistent with our
value of $\mathcal{F}=0.54\pm0.02$, but the fit column density was
$\nh=(1.0\pm0.1)\times10^{24}$\,\pcmsq, which is a factor of two
larger than what we found. The spectra are consistent with the
simple and reasonable physical picture that the cause of anomalous
and normal low states are the same, specifically occultation of
the \xray\ source by the accretion disk.

The \pca\ light curve (Fig.~\ref{fig:pca_lc}) shows a gradual drop
of flux by 40\% through the observation. We compared the spectra
from the four \rxte\ orbits and found that the spectral shape and
fractional RMS variation remained the same. This is inconsistent
with a change in the amount of cold absorber during the
observation, which would indicate an inhomogeneous screen of
material moving across the line of sight. Because of this, we
interpret the 30\% unabsorbed spectral component as scattering
into the line of sight by a hot corona rather than transmission
through a patchy, partially covering disk. A scattering corona of
the right size can also explain the change in the pulsations. The
fractional RMS variation in the \als\ is reduced by a factor of
\wsim14 from the main-on and the pulse itself is broader
(Fig.~\ref{fig:pca_flc}). A corona of moderate optical depth that
is smaller than \wsim1.2 light seconds (the pulse period) would
scatter some of the pulsed emission into the line of sight without
completely washing out the pulsations. The 40\% reduction in flux
can be explained as increased scattering or increased occultation
by the accretion disk. But a reduction in \xray\ production at the
source with orbital phase cannot be ruled out.

Finally, the binary orbital phase of the observation is 0.38-0.52,
too early for the 40\% decrease in flux during the observation to
be due to a pre-eclipse dip. Furthermore, dip light curves show a
variable absorption column with a time scale of 30 seconds
\citep{ste99,lea97,rey95}. There is no evidence for a variable
absorption column during the observation, but the decrease could
still be interpreted as an anomalous dip, a drop in the intensity
that occasionally occurs shortly after main-on turn-on
\citep{kus99}. In ordinary main-on states, the anomalous dips can
be interpreted as due to occultations of the central source by the
wobbling outer edge of the accretion disk \citep{sha99,cro80}. In
the model of \citet{sha99}, during the \als s the outer edge of
the disk continuously screens the central source due to its lower
inclination to the line of sight (see above), but tidally induced
wobbling should still be present. So the hot corona around the
inner parts of the accretion disk is expected to be screened by
the outer disk wobbling shortly after main-on turn-on, exactly in
the same manner as the anomalous dips are produced. This can take
place on the first orbit after the expected, but as of now
unobserved, turn-on. Anomalous dips of the \xrays\ scattered by
the hot corona should be smoother and shallower than ordinary
anomalous dips because it is an extended corona and not the
central source that is being occulted.

%% ********************************************************************
%% ** Summary                                                        **
%% ********************************************************************
\section{Summary}

Optical observations of HZ Her show that accretion is continuing
and \xrays\ are being produced at the neutron star at near the
normal level. The onset of the \als\ was accompanied by a
spin-down of the neutron star, a significant deviation from the
normal spin-up trend, which implies a slight drop in accretion
rate onto the neutron star. In the model of \citet{sha99} a
decrease in mass accretion would cause the inclination of the
accretion disk to lessen, and thus the outer edges of the
accretion disk could continuously screen the \xray\ source. Given
the caveat that our \rxte\ observation might not have been made
during an expected main-on (see Section 2), the \als\ is
spectrally very similar to the normal low state. Both low state
spectra can be fit with a partially absorbed, partially scattered
model where the input spectral shape is the same as in the
main-on. The \als, normal low state, and main-on have the same
intrinsic spectral shape, and no drastic change in the source
spectrum or luminosity is necessary. So the low state spectra
(anomalous and normal) are consistent with a single mechanism,
namely occultation of the neutron star by the accretion disk.  We
find that the combined optical, spectral, and timing observations
of the \als\ are consistent with a decrease in the inclination
angle of the disk, screening the \xray\ source over the entire
35\,d cycle, and a significant change in the warp of the disk is
not required.

\acknowledgments We thank E. Smith and J. Swank for rapidly
scheduling the observation.  This work was supported by NASA grant
NAS5-30720, NATO grant PST.CLG 975254, RFBR grant 98-02-16801, and
NSF Travel Grant NSF INT-9815741.

\clearpage

%% ********************************************************************
%% ** Bibliography                                                   **
%% ********************************************************************

\clearpage

%% ********************************************************************
%% ** Figures                                                        **
%% ********************************************************************

\clearpage

\clearpage

\clearpage

\clearpage

\clearpage

\clearpage

\begin{table}[ht]
\caption{\label{table:obs} \her\ Observations}
\begin{minipage}{\linewidth}
\renewcommand{\thefootnote}{\thempfootnote}
\begin{tabular}{lllcclcc} \hline \hline
 & & &\multicolumn{2}{c}{On-Source Livetime (ks)} & & \multicolumn{2}{c}{PCA} \\
 \cline{4-5} \cline{7-8}
 State & Date & Orbital Phase & PCA & HEXTE & &
 PCUs\footnote{Number of \pca\ detectors operating} &
 Epoch\footnote{\pca\ gain epoch} \\ \hline
 ALS & 1999 Apr. 26 & 0.38-0.52 & 4.6 & 2.8 & & 4 & 4 \\ \hline
 Main-On & 1997 Sep. 14 & 0.17-0.35 & 15.1 & 9.2 & & 5 & 3 \\
 Low & 1997 Sep. 12 & 0.11-0.21 & 3.3 & 2.0 & & 5 & 3 \\ \hline
\end{tabular}
\end{minipage}
\end{table}

\clearpage

\begin{table}[ht]
\caption{\label{table:fits} Fit spectral parameters for the \her\
main-on, anomalous low, and low states}
\begin{minipage}{\linewidth}
\renewcommand{\thefootnote}{\thempfootnote}
\begin{tabular}{llllll} \hline \hline
 Parameter
 & \label{col:mainon}Main-On\footnote{Fit with Eq.~\ref{eq:conteq}}
 &\label{col:als-a}ALS\footnotemark[\value{mpfootnote}]
 &\label{col:als-b}ALS\footnote{Fit with Eq.~\ref{eq:model} and the \hecut\
  shape free}
 &\label{col:als-c} ALS\footnote{Fit with Eq.~\ref{eq:model} and the main-on
  \hecut\ shape}
 &\label{col:low}Low\footnotemark[\value{mpfootnote}] \\ \hline

 $\nhu$ & 0.0\footnote{Not allowed to vary} & $< 0.4$ & $45.\pm6.$ & $50.\pm3.$ & $48.\pm4.$\\

 \wgamma & $1.08\pm0.01$ & $0.37\pm0.04$ & $0.90\pm0.07$ &
 1.08\footnotemark[\value{mpfootnote}] & 1.08\footnotemark[\value{mpfootnote}]\\

 $E_{c}$\,(keV) & $21.5\pm0.1$ & $16.8\pm0.4$ & $20.1\pm0.7$ &
 21.5\footnotemark[\value{mpfootnote}] & 21.5\footnotemark[\value{mpfootnote}]\\

 $E_{f}$\,(keV) & $9.8\pm0.1$ & $11.\pm1.$ & $10.\pm1.$ &
 9.8\footnotemark[\value{mpfootnote}] & 9.8\footnotemark[\value{mpfootnote}]\\

 FeK E\,(keV) & $6.48\pm0.09$ & $6.43\pm0.08$ & $6.44\pm0.08$ &
 6.48\footnotemark[\value{mpfootnote}] & 6.48\footnotemark[\value{mpfootnote}]\\

 FeK\ EW\,(keV) & $0.3\pm0.1$ & $0.7\pm0.1$ & $0.7\pm0.1$ & $0.7\pm0.1$ & $0.5\pm0.1$\\

 Flux\footnote{20-40\,keV flux in units of $10^{-10}$\,\flux} &
 $24.43\pm0.09$ & $1.50\pm0.07$ & $1.39\pm0.06$ & $1.41\pm0.06$ & $1.43\pm0.04$ \\

 $\cal F$ & N/A & N/A & $0.56\pm0.03$ & $0.67\pm0.04$ & $0.54\pm0.02$\\

 \wchired/DOF & 4.01/222\footnote{The high \wchired is a result of no
 CRSF in the fit (see text)} & 1.12/114 & 1.08/113 & 1.07/117 & 1.06/116\\ \hline
\end{tabular}
\end{minipage}
\end{table}


\begin{thebibliography}{}

\bibitem[Bildsten et al.(1997)]{bil97} Bildsten, L. et al., 1997,
\apjs, 113, 367

\bibitem[Brainerd \& Lamb(1987)]{bra87} Brainerd, J. \& Lamb, F.
K. 1987, \apj, 317, L33

\bibitem[Crosa \& Boynton(1980)]{cro80} Crosa, L., \& Boynton, P. E. 1980,
\apj, 235, 999

\bibitem[Dal Fiume et al.(1998)]{dal98} Dal Fiume, D., et al. 1998, \aap,
329, L41

\bibitem[Deeter et al.(1991)]{dee91} Deeter, J. E., et al. 1991, \apj, 383,
324

\bibitem[Deeter, Boynton, \& Pravdo(1981)]{dee81} Deeter, J. E., Boynton,
P. E., \& Pravdo, S. H. 1981, \apj, 247, 1003

\bibitem[Deeter et al.(1998)]{dee98} Deeter, J. E., Scott, D. M.,
Boynton, P. E., Miyamoto, S., Kitamoto, S., Takahama, S. \&
Fumiaki, N. 1998, \apj, 502, 802

\bibitem[Delgado, Schmidt, \& Thomas(1983)]{del83} Delgado, A. J., Schmidt,
H. U., \& Thomas, H. C. 1983, \aap, 127, L15

\bibitem[Doxsey et al.(1973)]{dox73} Doxsey, R., Bradt, H. V., Levine, A.,
Murthy, G. T., Rappaport, S., \& Spada, G. 1973, ApJ, 182, L25

\bibitem[Giacconi et al.(1973)]{gia73} Giacconi, R., Gursky, H., Kellogg,
E., Levinson, R., Schreier, E., \& Tananbaum, H. 1973 \apj, 184,
227

\bibitem[Ghosh \& Lamb(1979)]{gho79} Ghosh, P., \& Lamb, F. K. 1979, \apj,
234, 296

\bibitem[Gottwald et al.(1991)]{got91} Gottwald, M., Steinle, H., Graser,
U., \& Pietsch, W. 1991, \aaps, 89, 367

\bibitem[Gruber et al.(1998)]{gru98} Gruber, D. E., Heindl, W. A.,
Rothschild, R. E., Staubert, R., \& Wilms, J. 1998, Nuclear
Physics B (Proc. Suppl), 69, 174

\bibitem[Jahoda et al.(1996)]{jah96} Jahoda, K., Swank, J. H., Giles,
A. B., Stark, M. J., Strohmayer, T., \& Zhang, W. 1996, \procspie,
2808, 59

\bibitem[Kreykenbohm et al. (1999)]{kre99} Kreykenbohm, I., Kretschmar, P.,
Wilms, J., Staubert, R., Kendziorra, E., Gruber, D., Heindl, W.,
\& Rothschild, R., 1999, \aap, 341, 141

\bibitem[Kuster et al.(1999)]{kus99} Kuster, M., Wilms, J., Blum, S.,
Staubert, R., Gruber, D., Rothschild, R., \& Heindl, W. 1999, in
Proc. of the 3rd Integral Workshop, in press

\bibitem[Kylafis \& Klimis(1987)]{kyl87} Kylafis, N. D. \& Klimis, G. S.
1987, \apj, 323, 678

\bibitem[Leahy(1997)]{lea97} Leahy, D. A. 1997, \mnras, 287, 622

\bibitem[Levine(1999)]{lev99} Levine, A. M. 1999, \iaucirc, No. 7139

\bibitem[Lyutyi(1999)]{lyu99} Lyutyi, V. M., \& Goranskii, V. P.
2000, Astron. Lett., submitted

\bibitem[Mihara et al.(1991)]{mih91} Mihara, T., Ohashi, T., Makishma,
K., Nagase, F., Kitamoto, S., \& Koyama, K. 1991, \pasj, 43, 501

\bibitem[Mihara(1995)]{mih95} Mihara, T. 1995, Ph.D. Thesis,
University of Tokyo

\bibitem[Mironov et al.(1986)]{mir86} Mironov, A. V., Moshkalev, V. G.,
Trunkovskii, E. M., \& Cherepashchuk, A. M. 1986, SvA, 30, 68

\bibitem[Morrison \& McCammon(1983)]{mor83} Morrison, R., \& McCammon,
D. 1983, \apj, 270, 119

\bibitem[Parmar et al.(1985)]{par85} Parmar, A. N., Pietsch, W., McKechnie,
S., White, N. E., Tr\"{u}mper, J., Voges, W., \& Barr, P. 1985,
\nat, 313, 119

\bibitem[Parmar et al.(1999)]{par99} Parmar, A. N., Oosterbroek, T., Dal
Fiume, D., Orlandini, M., Santangelo, A., Segreto, A., \& Del
Sordo, S. 1999, \aap, 350, L5

\bibitem[Priedhorsky \& Holt(1987)]{pri87} Priedhorsky, W. C., \& Holt,
S. S. 1987, Space Sci. Rev. E, 45, 291

\bibitem[Pringle(1996)]{pri96} Pringle, J. E. 1996, \mnras, 281, 357

\bibitem[Reynolds \& Parmar(1995)]{rey95}
Reynolds, A. P. \& Parmar, A. N. 1995, \aap, 297, 747

\bibitem[Rothschild et al.(1998)]{rot98} Rothschild, R., et al. 1998, \apj,
496, 538

\bibitem[Shakura et al.(1998)]{sha98} Shakura, N. I., Ketsaris, N. A.,
Prokhorov, M. E., \& Postnov, K. A. 1998, \mnras, 300, 992

\bibitem[Shakura et al.(1999)]{sha99} Shakura, N. I., Prokhorov, M. E.,
Postnov, K. A., \& Ketsaris, N. A. 1999, \aap, 348, 917

\bibitem[Schandl(1996)]{sch96} Schandl, S. 1996, \aap, 307, 95

\bibitem[Schandl \& Meyer(1994)]{sch94} Schandl, S., \& Meyer., F. 1994,
\aap, 289, 149

\bibitem[Scott \& Leahy(1999)]{sco99} Scott, D. M., \& Leahy, D. A. 1999,
\apj, 510, 974

\bibitem[Scott et al.(2000)]{sco00} Scott, D. M., Leahy, D. A. \&
Wilson, R. B. 2000, submitted to \apj

\bibitem[Stelzer et al.(1999)]{ste99} Stelzer, B., Wilms, J., Staubert, R.,
Gruber, D., \& Rothschild, R. 1999, \aap, 342, 736

\bibitem[Tananbaum et al.(1972)]{tan72} Tananbaum, H., Gursky, H., Kellogg,
E. M., Levinson, R., Schreier, E., \& Giacconi, R. 1972, \apj,
174, L143

\bibitem[Vrtilek et al.(1994)]{vrt94} Vrtilek, S. D., et al. 1994, \apj,
436, L9

\bibitem[Vrtilek et al.(1996)]{vrt96} Vrtilek, S. D., \& Cheng, F. H. 1996,
\apj, 465,915

\end{thebibliography}
\end{document}